\documentclass[aps,pra,twocolumn,groupedaddress]{revtex4-2}

\usepackage[dvips]{graphicx}
\usepackage{braket}
\usepackage{amsmath}%
\DeclareMathOperator{\sign}{sgn}

\begin{document}
\renewcommand{\vec}[1]{\mathbf{#1}}
% You should use BibTeX and apsrev.bst for references
% Choosing a journal automatically selects the correct APS
% BibTeX style file (bst file), so only uncomment the line
% below if necessary.
\bibliographystyle{apsrev4-2}

%\preprint{}

\title{On the diagonalization of quadratic Hamiltonians}
\author{Ville J. H\"{a}rk\"{o}nen}
\email[]{ville.j.harkonen@gmail.com}
\affiliation{Department of Applied Physics, Aalto University School of Science, FI-00076 Aalto, Finland}

\author{Ivan A. Gonoskov}
\email[]{ivan.gonoskov@gmail.com}
\affiliation{Institute for Physical Chemistry, Friedrich-Schiller-University Jena, Max-Wien-Platz 1, 07743 Jena, Germany}
%\homepage[]{https://www.jyu.fi/kemia/en/research/main-group-chemistry/personnel/vharkonen}
%\thanks{}
\date{\today}

\begin{abstract}
A new procedure to diagonalize quadratic Hamiltonians is introduced. We show that one can find a unitary transformation such that the transformed quadratic Hamiltonian is diagonal but still written in terms of the original position and momentum observables. We give a general method to diagonalize an arbitrary quadratic Hamiltonian and derive a few of the simplest special cases in detail.
\end{abstract}


\maketitle


\section{Introduction}
\label{cha:Introduction}

Quadratic Hamiltonian plays a special role in the history of quantum physics. Not only being a famous solvable example at the dawn of quantum mechanics, it is exact physical Hamiltonian for such fundamental quantum systems as a quantized electromagnetic field in a vacuum and a free electron in a uniform magnetic field. The range of validity of the quadratic Hamiltonian is vast. For instance, the Hamiltonians of this form appear whenever one studies the nuclear contribution in the Born-Oppenheimer (BO) approximation \cite{Born-Oppenheimer-Adiabatic-Approx.1927,born-huang-dynamical-1954} and in the description of nuclear dynamics beyond the BO approximation \cite{Requist-ExactFactorizationBasedDFTofElectronPhononSystems-PhysRevB.99.165136-2019,Harkonen-ManybodyGreensFunctionTheoryOfElectronsAndNucleiBeyondTheBornOppenheimerApproximation-PhysRevB.101.235153-2020}. Therefore these Hamiltonians play a central role in the study of phonon spectrum in crystals \cite{Giannozzi-AbInitioCalcOfPhononDispersInSemicond-PhysRevB.43.7231-1991,Baroni-PhononsAndRelatedCrystalPropFromDFTPT-RevModPhys.73.515-2001,Harkonen-NTE-2014,Togo-FirstPrincPhononCalcInMaterScience-2015,Ribeiro-StrongAnharmonInThePhononSpectraOfPbTeAndSnTeFromFirstPrinc-PhysRevB.97.014306-2018}, conventional superconductivity \cite{Errea-HighPressureHydrogenSulfideFromFirstPrinciplesAStronglyAnharmonicPhononMediatedSuperconductor-PhysRevLett.114.157004-2015,Sun-RouteToASuperconductingPhaseAboveRoomTemperatureInElectronDopedHydrideCompoundsUnderHighPressure-PhysRevLett.123.097001-2019,Somayazulu-EvidenceForSuperconductivityAbove260KInLanthanumSuperhydrideAtMegabarPressures-PhysRevLett.122.027001-2019}, thermal conductivities \cite{Harkonen-Tcond-II-VIII-PhysRevB.93.024307-2016,Harkonen-AbInitioComputStudyOnTheLattThermalCondOfZintlClathrates-PhysRevB.94.054310-2016,Feng-FourPhononScatteringSignificantlyReducesIntrinsicThermalConductivityOfSolids-PhysRevB.96.161201-2017} and molecular vibrations \cite{Wilson-MolecularVibrationsTheTheoryOfInfraredAndRamanVibrationalSpectra-1955}. Further, these Hamiltonians are also relevant in the study of quantum entanglement \cite{Audenaert-EntanglementPropertiesOfTheHarmonicChain-PhysRevA.66.042327-2002,Plenio-DynamicsAndManipulationOfEntanglementInCoupledHarmonicSystemsWithManyDegreesOfDreedom-2004,Adesso-MultipartiteEntanglementInThreeModeGaussianStatesOfContinuousVariableSystemsQuantificationSharingStructureAndDecoherence-PhysRevA.73.032345-2006,Anders-EntanglementAndSeparabilityOfQuantumHarmonicOscillatorSystemsAtFiniteTemperature-2007,Anders-ThermalStateEntanglementInHarmonicLattices-PhysRevA.77.062102-2008,Makarov-CoupledHarmonicOscillatorsAndTheirQuantumEntanglement-PhysRevE.97.042203-2018,Park-DynamicsOfEntanglementAndUncertaintyRelationInCoupledHarmonicOscillatorSystemExactResults-2018,Park-DynamicsOfEntanglInThreeCoupledHarmonicOscillatorSystemWithArbitraryTimeDependentFrequencyAndCouplingConstants-2019,Merdaci-EntanglementInThreeCoupledHarmonicOscillators-2020,Lydzba-EigenstateEntanglementEntropyInRandomQuadraticHamiltonians-PhysRevLett.125.180604-2020,Lydzba-EntanglementInManyBodyEigenstatesOfQuantumChaoticQuadraticHamiltonians-PhysRevB.103.104206-2021}, in the field of quantum optics \cite{Scully-QuantumOptics-1997} and in the analysis of light-matter interaction within full quantum description \cite{Gonoskov-QuantumOpticalSignaturesInStrongFieldLaserPhysicsInfraredPhotonCountingInHighOrderHarmonicGeneration-2016,Gonoskov-LightMatterQuantumDynamicsOfComplexLaserDrivenSystems-2021}.

The exact diagonalization of the quadratic Hamiltonian is a standard procedure which can be done with different approaches. From these available approaches, one can choose the most suitable one for the problem at hand. For instance, the phonon representation is useful in the study on lattice dynamic related problems \cite{born-huang-dynamical-1954} and normal coordinates in the study of molecular vibrations \cite{Wilson-MolecularVibrationsTheTheoryOfInfraredAndRamanVibrationalSpectra-1955}. Yet another way is to use symplectic transformations \cite{Dutta-TheRealSymplecticGroupsInQuantumMechanicsAndOptics-1995,Han-IllustrativeExampleOfFeynmansRestOfTheUniverse-1999} to establish the diagonalization which has turned out to be useful in the study of entanglement \cite{Eisert-IntroductionToTheBasicsOfEntanglementTheoryInContinuousVariableSystems-2003,Cramer-CorrelationsSpectralGapAndEntanglementInHarmonicQuantumSystemsOnGenericLattices-2006,Makarov-CoupledHarmonicOscillatorsAndTheirQuantumEntanglement-PhysRevE.97.042203-2018}. What unites these approaches is that after the transformation the Hamiltonian is diagonal in the new coordinates which in turn are some collective coordinates of the original position and momentum variables.

In this work, we take yet another root and diagonalize the quadratic Hamiltonian by changing the quantum reference frame \cite{Aharonov-QuantumFramesOfReference-PhysRevD.30.368-1984,Giacomini-QuantumMechanicsAndTheCovarianceOfPhysicalLawsInQuantumReferenceFrames-2019} by a unitary transformation. Similar transformations were introduced in Ref. \cite{Gonoskov-IonizationInaQuantizedElectromagneticField-2007}. The transformed Hamiltonian is diagonal, but this time in the original position and momentum variables the difference of the original and transformed Hamiltonians being in the masses and coupling constants.

This paper is organized as follows. In Sec. \ref{cha:QuadraticHamiltonians}, the generic quadratic Hamiltonian is given, we discuss some of its properties and how the diagonalization is usually established. We introduce a transformation to diagonalize generic $n$-body quadratic Hamiltonians in Sec. \ref{cha:DiagonalizationProcedure} and discuss the physical interpretation of our results. A special case of $2$-body system is discussed in Sec \ref{cha:TwoBodyCase} and one dimensional chain with nearest neighbour interactions in Sec. \ref{cha:ChainWithNearestNeighbourInteraction}. We solve the $3$-body case in Appendix \ref{ThreeBodyCase}.

\section{Quadratic Hamiltonians}
\label{cha:QuadraticHamiltonians}

The object of our study is the quadratic Hamiltonian of the form
\begin{equation} 
H = \sum^{n}_{i} \frac{ p^{2}_{i} }{ 2 m_{i} } + \frac{1}{2}\sum^{n}_{i,j} \Phi_{ij} u_{i} u_{j},
\label{eq:QuadraticHamiltoniansEq_1}
\end{equation}
where $m_{i}$ is the mass of the $i$th particle, $u_{i}$ the position operator of the particle $i$ and $p_{i} = -i \hbar \partial/\partial{u_{i}}$ the corresponding momentum. The quantities $\Phi_{ij}$ are the coupling constants and will be considered as parameters. For example, in the BO context, the quantities $\Phi_{ij}$ are the so-called interatomic force constants, the second-order derivatives of the BO energy surface with respect to the nuclear equilibrium positions. There are many alternative approaches to diagonalize quadratic Hamiltonians like $H$ and here we provide a summary of them. By doing so, we obtain some central results to be compared with those obtained in this work.

From the structure of the Hamiltonian $H$ it follows that we can write the potential part in terms of symmetric parameters. Sometimes the parameters $\Phi_{ij}$ are symmetric themselves. For such symmetric matrices, the diagonalization can be established by an orthogonal coordinate transformation. For the sake of clarity we summarize the diagonalization procedure \cite{born-huang-dynamical-1954}. We assume that the parameters are symmetric $\Phi_{ij} = \Phi_{ji}$. The symmetric matrix $D_{ij} \equiv \Phi_{ij} / \sqrt{m_{i}m_{j}}$ satisfies
\begin{equation} 
\omega^{2}_{s} e_{is} = \sum^{n}_{j} D_{ij} e_{js}, \quad  s = 1,\ldots,n,
\label{eq:QuadraticHamiltoniansEq_2}
\end{equation}
where the eigenvectors have the following properties
\begin{equation}
\sum_{i} e_{is'} e_{is} = \delta_{ss'}, \quad \sum_{s} e_{is} e_{i's} = \delta_{ii'}.
\label{eq:QuadraticHamiltoniansEq_3}
\end{equation}
We transform the Hamiltonian to mass scaled coordinates $w_{i} \equiv \sqrt{m_{i}} u_{i}$ and then transform these coordinates with the eigenvectors as
\begin{equation}
w_{i} = \sum_{s} e_{is} q_{s}.
\label{eq:QuadraticHamiltoniansEq_4}
\end{equation}
After establishing the corresponding transformations for momentum we obtain the diagonal form of the Hamiltonian
\begin{equation}
H = \frac{ 1 }{ 2 } \sum^{n}_{ s = 1 } \left( - \hbar^{2}  \frac{\partial^{2}{ }}{ \partial{ q^{2}_{s} } } + \omega^{2}_{s}   q^{2}_{s} \right).
\label{eq:QuadraticHamiltoniansEq_5}
\end{equation}
We can further write Eq. \ref{eq:QuadraticHamiltoniansEq_5} in terms of creation and annihilation operators as
\begin{equation} 
H = \sum^{n}_{ s = 1 } \hbar \omega_{s} \left( \frac{1}{2} + a^{\dagger}_{s} a_{s}\right).
\label{eq:QuadraticHamiltoniansEq_6}
\end{equation}
Here, the creation operator is defined as $a^{\dagger}_{s} \equiv \left( \omega_{s} q_{s} - i p_{s} \right) / \sqrt{2 \hbar \omega_{s} }$ and $p_{s} = -i \hbar \partial/\partial{q_{s}}$. The interpretation of $a^{\dagger}_{s}$ is that it creates a quantum of energy $\hbar \omega_{s}$ on the vibrational mode $s$. This vibrational mode comprises a collective motion of several particles described by the coordinates $u_{i}$ since the coordinates $q_{s}$ are collective. This collective motion is due to the non-diagonal interaction terms in the Hamiltonian and the strength of the interaction is determined by the parameters $\Phi_{ij}$. If $\Phi_{ij} = 0$ for all $i \neq j$, the Hamiltonian given by Eq. \ref{eq:QuadraticHamiltoniansEq_1} describes independent harmonic oscillators with the frequencies $\omega^{2}_{i} = \Phi_{ii} / m_{i}$. We refer to the frequency $\omega_{i}$ as the frequency of the $i$th independent harmonic oscillator. With the non-diagonal terms vanishing, each of the particle coordinates $u_{i}$ are independent and the total wave function is of the product form. These functions for each $u_{i}$ are of the simple harmonic oscillator form.

Another commonly used approach to diagonalize $H$ is the use of phonon coordinates \cite{born-huang-dynamical-1954,maradudin-harm-appr-1971}. The procedure comprises establishing the transformation to mass scaled coordinates, Fourier transforming the mass scaled coordinates (periodic boundary conditions used) and finally transforming to the phonon coordinates by using the eigenvectors of the dynamical matrix. The dynamical matrix is the Fourier transform of the interatomic force constant matrix ($D_{ij}$ in our notation). After this procedure, the Hamiltonian is again diagonal and can be written in terms of creation and annihilation operators as in Eq. \eqref{eq:QuadraticHamiltoniansEq_6}, but this time the collective coordinates are different. However, the principle and the physical picture remains similar and in the excitation of a phonon mode, a collective motion of the original particle coordinates occurs.

The approach \cite{Dutta-TheRealSymplecticGroupsInQuantumMechanicsAndOptics-1995,Han-IllustrativeExampleOfFeynmansRestOfTheUniverse-1999} used in the field of entanglement \cite{Eisert-IntroductionToTheBasicsOfEntanglementTheoryInContinuousVariableSystems-2003,Cramer-CorrelationsSpectralGapAndEntanglementInHarmonicQuantumSystemsOnGenericLattices-2006,Makarov-CoupledHarmonicOscillatorsAndTheirQuantumEntanglement-PhysRevE.97.042203-2018} does not differ from the afore mentioned approaches in that the diagonal form of the Hamiltonian is obtained in the transformed coordinates. The transformed coordinates, in turn, involve two or more of the original position observables.

\section{New diagonalization procedure}
\label{cha:DiagonalizationProcedure}

Here we introduce a new transformation to diagonalize $H$. The starting point is the time-independent Schr\"{o}dinger equation
\begin{equation} 
H \tilde{\chi} = E\tilde{\chi}.
\label{eq:NewWayOfDiagonalizationEq_1}
\end{equation}
We assume that $U \chi =  \tilde{\chi}$, where $U$ is some suitable transformation, an explicit form of which will be given later. We write the corresponding Schr\"{o}dinger equation for the transformed system as
\begin{equation} 
\tilde{H} \chi = E \chi, \quad \tilde{H} = U^{-1} H U.
\label{eq:NewWayOfDiagonalizationEq_2}
\end{equation}
We seek a transformation $U$ such that $\tilde{H}$ is diagonal. Consider a transformation of the form
\begin{equation} 
U \equiv e^{ \alpha u_{1} \frac{\partial}{\partial{u_{2}}} }, \quad U^{-1} = e^{ -\alpha u_{1} \frac{\partial}{\partial{u_{2}}} },
\label{eq:NewWayOfDiagonalizationEq_3}
\end{equation}
where $\alpha$ is some real parameter. We see that the operator $u_{1} \partial/\partial{u_{2}} = i u_{1} p_{2} / \hbar$ is Hermitian and therefore $U$ is unitary, $U^{-1} = U^{\dagger}$. The transformation $U$ is a translation operator such that when we act on a function $f\left(u_{1},u_{2}\right)$ it follows that
\begin{eqnarray} 
U f\left(u_{1},u_{2}\right) &=& f\left(u_{1}, u_{2} + \alpha u_{1} \right), \nonumber \\
U^{\dagger} f\left(u_{1},u_{2}\right) &=& f\left(u_{1}, u_{2} - \alpha u_{1} \right).
\label{eq:NewWayOfDiagonalizationEq_4}
\end{eqnarray}
With Eqs. \ref{eq:NewWayOfDiagonalizationEq_3} and \ref{eq:NewWayOfDiagonalizationEq_4} we can show that
\begin{eqnarray} 
U^{\dagger} u_{1} U &=& u_{1}, \quad U^{\dagger} u_{2} U = u_{2} - \alpha u_{1}, \nonumber \\
U^{\dagger} p_{2} U &=& p_{2}, \quad U^{\dagger} p_{1} U = \alpha p_{2} + p_{1}.
\label{eq:NewWayOfDiagonalizationEq_5}
\end{eqnarray}
Given these results we see that the quadratic Hamiltonian $H$ is still quadratic after the transformation. That is, $\tilde{H}$ is quadratic after using $U$ given by Eq. \ref{eq:NewWayOfDiagonalizationEq_3}. However, the Hamiltonian is not necessary diagonal and in general non-diagonal terms appear also in the kinetic energy. We claim that we can find a suitable form of $U$ and parameters like $\alpha$ such that the resulting quadratic Hamiltonian $\tilde{H}$ is diagonal. With these results in mind, we formulate our new diagonalization procedure in the general case as follows.

We re-write the Hamiltonian of Eq. \ref{eq:QuadraticHamiltoniansEq_1} as
\begin{equation} 
H = \sum^{n}_{i} \frac{ p^{2}_{i} }{ 2 m_{i} } + \sum^{n}_{i} d_{i} u^{2}_{i} + \sum^{n}_{i<j} d_{ij} u_{i} u_{j},
\label{eq:NewWayOfDiagonalizationEq_6}
\end{equation}
where
\begin{equation} 
d_{i} \equiv \frac{1}{2} \Phi_{ii}, \quad d_{ij} \equiv \frac{1}{2} \left( \Phi_{ij} + \Phi_{ji} \right).
\label{eq:NewWayOfDiagonalizationEq_7}
\end{equation}
We would like to find a transformation which acts on all the cross terms with the parameters $d_{ij}$ separately and thus the number of these transformations is $1/2 \left(n-1\right) n \equiv N$. We choose the total transformation as
\begin{eqnarray} 
U &\equiv&  U_{1} \cdots U_{N}, \nonumber \\
U_{1} &=& \text{exp}\left[ \alpha_{1} u_{2} \frac{\partial{ }}{\partial{u_{1}}} \right] \text{exp}\left[ \beta_{1} u_{1} \frac{\partial{ }}{\partial{u_{2}}} \right], \nonumber \\
U_{2} &=& \text{exp}\left[ \alpha_{2} u_{3} \frac{\partial{ }}{\partial{u_{1}}} \right] \text{exp}\left[ \beta_{2} u_{1} \frac{\partial{ }}{\partial{u_{3}}} \right], \nonumber \\
      &\vdots& \nonumber \\
U_{N} &=& \text{exp}\left[ \alpha_{N} u_{n} \frac{\partial{ }}{\partial{u_{n-1}}} \right] \text{exp}\left[ \beta_{N} u_{n-1} \frac{\partial{ }}{\partial{u_{n}}} \right],
\label{eq:NewWayOfDiagonalizationEq_8}
\end{eqnarray}
where
\begin{eqnarray} 
\beta_{1} &=& - \frac{  \alpha_{1} }{ \alpha^{2}_{1} + \frac{m_{2}}{ m_{1} } }, \nonumber \\
\beta_{2} &=& - \frac{  \alpha_{2} }{ \alpha^{2}_{2} + \frac{m_{3}}{ m_{1} } }, \nonumber \\
      &\vdots& \nonumber \\
\beta_{N} &=& - \frac{  \alpha_{N} }{ \alpha^{2}_{N} + \frac{m_{n}}{ m_{n-1} } }.
\label{eq:NewWayOfDiagonalizationEq_9}
\end{eqnarray}
After each of the transformations, $U_{k}$, the Hamiltonian $\tilde{H}$ remains quadratic. After the transformation, the Hamiltonian can be written as
\begin{equation} 
\tilde{H} = \sum^{n}_{i} \frac{ p^{2}_{i} }{ 2 \tilde{m}_{i} } + \sum^{n}_{i,j} T_{ij} p_{i} p_{j} + \sum^{n}_{i} \tilde{d}_{i} u^{2}_{i} + \sum^{n}_{i<j} \tilde{d}_{ij} u_{i} u_{j}.
\label{eq:NewWayOfDiagonalizationEq_10}
\end{equation}
The aim is to find the parameters  $\alpha_{k},\beta_{k}$ such that
\begin{equation} 
T_{ij} = 0, \quad  \tilde{d}_{ij} = 0,
\label{eq:NewWayOfDiagonalizationEq_11}
\end{equation}
for all $i,j$. It turns out that $T_{ij} = 0$ is automatically satisfied when we use $\beta_{k}$ given by Eq. \ref{eq:NewWayOfDiagonalizationEq_9}, we see this when we go through the $2$-body case explicitly in Sec. \ref{cha:TwoBodyCase}. What is left is to solve the $N$ equations $\tilde{d}_{ij} = 0$ such that each $\alpha_{k}$ can be written as a function of the original parameters $m_{i}$, $d_{i}$ and $d_{ij}$. If Eq. \ref{eq:NewWayOfDiagonalizationEq_11} holds, the transformed Hamiltonian reads
\begin{equation} 
\tilde{H} = \sum^{n}_{i} \left( \frac{ p^{2}_{i} }{ 2 \tilde{m}_{i} } + \tilde{d}_{i} u^{2}_{i} \right),
\label{eq:NewWayOfDiagonalizationEq_12}
\end{equation}
and we have reached our goal. We can further write Eq. \ref{eq:NewWayOfDiagonalizationEq_12} in terms of creation and annihilation operators as
\begin{equation} 
\tilde{H} = \sum^{n}_{i = 1} \hbar \omega_{i} \left( \frac{1}{2} + a^{\dagger}_{i} a_{i} \right),
\label{eq:NewWayOfDiagonalizationEq_13}
\end{equation}
where $\omega^{2}_{i} \equiv 2 \tilde{d}_{i}/\tilde{m}_{i}$ and the creation operator is defined as $a^{\dagger}_{i} \equiv \left( \tilde{m}_{i} \omega_{i} u_{i} - i p_{i} \right) / \sqrt{2 \tilde{m}_{i} \hbar \omega_{i} }$. These operators satisfy
\begin{equation} 
\left[a_{i}, a^{\dagger}_{j} \right]_{-} = \delta_{ij}, \quad \left[a_{i}, a_{j} \right]_{-} = \left[a^{\dagger}_{i}, a^{\dagger}_{j} \right]_{-} = 0,
\label{eq:NewWayOfDiagonalizationEq_15}
\end{equation}
and the physical interpretation of these operators is the usual one \cite{Dirac-PrinciplesOfQM-1958}.

The resulting Hamiltonian $\tilde{H}$ is diagonal and all the original interactions are hidden in the masses $\tilde{m}_{i}$ and coefficients $\tilde{d}_{i}$, which we call the effective masses and the effective force constants. Therefore a general quadratic Hamiltonian $H$ can be transformed to a diagonal form such that the resulting Hamiltonian $\tilde{H}$ seems to be the one of $n$ independent harmonic oscillators. We note that the position $u_{i}$ and momentum operators $p_{i}$ in $\tilde{H}$ are still the original observables, not the collective ones as in the case of conventional techniques in diagonalizing the quadratic Hamiltonian. The system seems to have only the non-interacting independent harmonic oscillators in terms of the original position observables, even though we are still working with the exact Hamiltonian. The independent harmonic oscillators are, however, different from those we obtain from the original Hamiltonian due to the effective masses $\tilde{m}_{i}$ and coefficients $\tilde{d}_{i}$ leading to the effective frequencies $\omega_{i}$. We have thus moved to a reference frame \cite{Aharonov-QuantumFramesOfReference-PhysRevD.30.368-1984,Giacomini-QuantumMechanicsAndTheCovarianceOfPhysicalLawsInQuantumReferenceFrames-2019} in which the original observables are decoupled.

The solution of the corresponding Schr\"{o}dinger equation is known and for instance the ground state wave function $\chi_{0}\left(u\right)$ satisfying $\tilde{H} \chi_{0} = E_{0} \chi_{0}$ (Eq. \ref{eq:NewWayOfDiagonalizationEq_2} for the ground state) can be written in terms of the functions
\begin{equation} 
\chi_{0}\left(u_{i}\right) = \left(\frac{ \tilde{m}_{i} \omega_{i} }{\pi \hbar } \right)^{1/4} \text{exp}\left[ - \frac{ \tilde{m}_{i} \omega_{i}  }{ 2 \hbar } u^{2}_{i} \right],
\label{eq:NewWayOfDiagonalizationEq_16}
\end{equation}
such that $\chi_{0}\left(u\right) = \chi_{0}\left(u_{1}\right) \cdots \chi_{0}\left(u_{n}\right)$. The total wave function is thus a product of the single particle functions since $\tilde{H}$ is diagonal. This means, by definition, that the wave function $\chi$ satisfying Eq. \ref{eq:NewWayOfDiagonalizationEq_2} is not entangled \cite{Amico-EntanglementInManyBodySystems-RevModPhys.80.517-2008}. On the other hand, the wave function for the Hamiltonian $H$ with cross terms is entangled from which we deduce that the transformation $U^{\dagger}$ disentangles the entangled wave function $\tilde{\chi}$ when it acts on it, namely $\chi = U^{\dagger} \tilde{\chi}$. Therefore, $U^{\dagger}$ is a disentangling transformation of the quadratic Hamiltonian wave functions satisfying Eq. \ref{eq:NewWayOfDiagonalizationEq_1}. The explicit form of such disentangling transformation can be seen by looking Eqs. \ref{eq:NewWayOfDiagonalizationEq_8} and \ref{eq:NewWayOfDiagonalizationEq_9}.

Here we described our new and general procedure to diagonalize quadratic Hamiltonians. Whether or not we are able to diagonalize the Hamiltonian with $U$ given by Eq. \ref{eq:NewWayOfDiagonalizationEq_8} depends on the fact, whether or not we are able to solve Eq. \ref{eq:NewWayOfDiagonalizationEq_11}. To answer this question, we discuss the $2$-body special case in Sec. \ref{cha:TwoBodyCase} and the special $n$-body case with the nearest neighbour interactions in Sec. \ref{cha:ChainWithNearestNeighbourInteraction}. We consider the 3-body case in Appendix \ref{ThreeBodyCase}.

\section{Special cases}
\label{cha:SpecialCases}

\subsection{Two-body case}
\label{cha:TwoBodyCase}

Here we consider the 2-body case by setting $n = 2$ in Eq. \ref{eq:NewWayOfDiagonalizationEq_6}. We diagonalize $H$ by transformation of the form (see Eq. \ref{eq:NewWayOfDiagonalizationEq_8})
\begin{equation} 
U = e^{ \alpha u_{2} \frac{\partial{ }}{\partial{u_{1}}} } e^{ \beta u_{1} \frac{\partial{ }}{\partial{u_{2}}} }.
\label{eq:TwoBodyCaseEq_1}
\end{equation}
We find that with this choice of $U$, the position operators $u_{i}$ and the corresponding momentum operators $p_{i}$ transform as
\begin{eqnarray} 
U^{\dagger} u_{1} U &=& u_{1} \left( 1 + \alpha \beta \right) - \alpha u_{2}, \nonumber \\
U^{\dagger} u_{2} U &=& u_{2} - \beta u_{1}, \nonumber \\
U^{\dagger} p_{1} U &=&  \beta p_{2} + p_{1}, \nonumber \\
U^{\dagger} p_{2} U &=& \left( 1 + \alpha \beta \right) p_{2} + \alpha p_{1},
\label{eq:TwoBodyCaseEq_2}
\end{eqnarray}
and the remaining relations can be found, for instance, by using $U^{\dagger} u_{i} u_{j}  U = U^{\dagger} u_{i} U U^{\dagger} u_{j} U$ and then Eq. \ref{eq:TwoBodyCaseEq_2}. By using these results, we find that the condition $T_{ij} = 0$ (Eq. \ref{eq:NewWayOfDiagonalizationEq_11}) in the present case reads
\begin{eqnarray} 
0 &=& \frac{ \beta }{ m_{1} }  + \frac{  \left( 1 + \alpha \beta \right) \alpha  }{ m_{2} } \Leftrightarrow  \nonumber \\
\beta &=& -\frac{  \alpha  }{ \frac{ m_{2} }{ m_{1} }  + \alpha^{2} },
\label{eq:TwoBodyCaseEq_3}
\end{eqnarray}
and therefore, with this choise of $\beta$, the kinetic energy is diagonal. Since the general transformation \ref{eq:NewWayOfDiagonalizationEq_8} is product of transformations \ref{eq:TwoBodyCaseEq_1}, the kinetic energy is diagonal after transforming the Hamiltonian of Eq. \ref{eq:NewWayOfDiagonalizationEq_6} by $U$ given by Eq. \ref{eq:NewWayOfDiagonalizationEq_8}. The condition $\tilde{d}_{ij} = 0$ of Eq. \ref{eq:NewWayOfDiagonalizationEq_11} can be written in the present case as
\begin{eqnarray} 
0 &=& d_{12} \left( 1  + 2 \alpha \beta \right) - 2 d_{1} \alpha  \left( 1 + \alpha \beta \right) - 2 d_{2} \beta \Leftrightarrow \nonumber \\
0 &=& a \alpha^{2} + b \alpha + c,
\label{eq:TwoBodyCaseEq_4}
\end{eqnarray}
where (we use $k_{ij} \equiv m_{i} / m_{j}$) 
\begin{equation} 
a \equiv - d_{12}, \quad b \equiv - 2 \left( d_{1} k_{21}  - d_{22} \right), \quad c \equiv d_{12} k_{21}.
\label{eq:TwoBodyCaseEq_5}
\end{equation}
We solve Eq. \ref{eq:TwoBodyCaseEq_4} for $\alpha$ and thus
\begin{equation} 
\alpha = - \frac{  d_{1} k_{21}  - d_{2} \pm \sqrt{  \left( d_{1} k_{21}  - d_{2} \right)^{2} + d^{2}_{12} k_{21} }  }{  d_{12} }.
\label{eq:TwoBodyCaseEq_6}
\end{equation}
Either of the solutions of Eq. \ref{eq:TwoBodyCaseEq_6} with $\pm$ for $\alpha$ is acceptable. We have now obtained the diagonal form of the Hamiltonian $\tilde{H}$ given by Eq. \ref{eq:NewWayOfDiagonalizationEq_12} with $n = 2$ and the corresponding effective masses and coefficients can be written as
\begin{eqnarray} 
\tilde{m}_{1} &\equiv& \frac{ m_{1} m_{2} }{ m_{2} + m_{1} \alpha^{2} }, \quad \tilde{m}_{2} \equiv m_{2}  + m_{1} \alpha^{2}, \nonumber \\
\tilde{d}_{1} &\equiv&  \frac{ d_{1} + d_{12} k_{12} \alpha + d_{2} k^{2}_{12} \alpha^{2} }{ \left( 1  + k_{12} \alpha^{2} \right)^{2} },  \nonumber \\
\tilde{d}_{2} &\equiv&  d_{1} \alpha^{2} +  d_{2} - d_{12} \alpha.
\label{eq:TwoBodyCaseEq_7}
\end{eqnarray}

We found that the system of two coupled harmonic oscillators can be considered as two independent ($n = 2$) harmonic oscillators still written in terms of the original operators of position and momentum, but with the effective masses $\tilde{m}_{i}$ and force constants $\tilde{d}_{i}$. We can find the exact wave function satisfying the corresponding Schr\"{o}dinger equation written for the Hamiltonian $H$ by using $\chi = U^{\dagger} \tilde{\chi}$. For instance, the ground state function can be written as $\tilde{\chi}_{0}\left(u\right) = \chi_{0}\left( u'_{1} \right) \chi_{0}\left( u'_{2} \right)$, where $u'_{1} = u_{1} + \alpha u_{2}$ and $u'_{2} =  u_{2} + \beta \left( u_{1} + \alpha u_{2} \right)$. The explicit form of the entangled $\tilde{\chi}_{0}\left(u\right)$ can be therefore found by using these results together with Eq. \ref{eq:NewWayOfDiagonalizationEq_16}.

\subsection{Chain with nearest neighbour interaction}
\label{cha:ChainWithNearestNeighbourInteraction}

In this section we consider a special case of the $n$-body Hamiltonian given by Eq. \ref{eq:NewWayOfDiagonalizationEq_6} with the nearest neighbour interactions. We can think of the situation as follows. Consider a linear chain of $n$ particles with one spatial dimension only. Suppose that the particles are arranged such that the observable $u_{1}$ is for the particle which is in the vicinity of the left most site, $u_{2}$ to the right hand side of this site and so on. The right most particle site is described with the observable $u_{n}$. If some periodic boundary conditions are imposed, then $u_{1}$ and $u_{n}$ could be observables of neighbouring sites. We further assume that only the nearest neighbours interact with each other, that is, in Eq. \ref{eq:NewWayOfDiagonalizationEq_6}  $\tilde{d}_{ij} = 0$ if $\left|i-j\right| > 1$. With this assumption we write Eq. \eqref{eq:NewWayOfDiagonalizationEq_6} as
\begin{equation} 
H = \sum^{n}_{i} \frac{ p^{2}_{i} }{ 2 m_{i} } + \sum^{n}_{i} d_{i} u^{2}_{i} + \sum^{n-1}_{i} d_{i\left(i+1\right)} u_{i} u_{i+1}.
\label{eq:ChainWithNearestNeighbourInteractionEq_1}
\end{equation}
In order to make use of our earlier results obtained for the two-body case we rearrange the Hamiltonian and write
\begin{eqnarray} 
H &=& \frac{ p^{2}_{1} }{ 2 m_{1} } + \frac{ p^{2}_{2} }{ 2 m_{2} } + d_{1} u^{2}_{1} + d_{2} u^{2}_{2} + d_{12} u_{1} u_{2} + \cdots \nonumber \\
&&+ \frac{ p^{2}_{n-1} }{ 2 m_{n-1} } + \frac{ p^{2}_{n} }{ 2 m_{n} } + d_{n-1} u^{2}_{n-1} + d_{n} u^{2}_{n} \nonumber \\
&&+ d_{\left(n-1\right)n} u_{n-1} u_{n}.
\label{eq:ChainWithNearestNeighbourInteractionEq_2}
\end{eqnarray}
We assume that $n$ is an even number. By doing so we do not miss any relevant properties of the system provided $n$ is sufficiently large. It can be seen that in the present case, the $n$-body Hamiltonian can be written as a sum of $2$-body Hamiltonians which we have already diagonalized in Sec. \ref{cha:TwoBodyCase}. That is, we need $n/2$ transformations to diagonalize this particular $n$-body Hamiltonian and we already know the resulting Hamiltonian and the quantities related from our previous results. Namely, the resulting Hamiltonian is given by Eq. \ref{eq:NewWayOfDiagonalizationEq_12} and it can be obtained with the following transformations and the other quantities involved
\begin{eqnarray} 
U_{i} &\equiv& \text{exp}\left[ \alpha_{i} u_{i} \frac{\partial{ }}{\partial{u_{i-1}}} \right] \text{exp}\left[ - \frac{  \alpha_{i} }{ \alpha^{2}_{i} + \frac{m_{i}}{ m_{i-1} } } u_{i-1} \frac{\partial{ }}{\partial{u_{i}}} \right], \nonumber \\
\tilde{m}_{i-1} &=& \frac{ m_{i-1} m_{i} }{ m_{i} + m_{i-1} \alpha^{2}_{i} }, \quad \tilde{m}_{i} = m_{i}  + m_{i-1} \alpha^{2}_{i}, \nonumber \\
\tilde{d}_{i-1} &=&  \frac{ d_{i-1} + d_{\left(i-1\right)i} k_{\left(i-1\right)i} \alpha_{i} + d_{i} k^{2}_{\left(i-1\right)i} \alpha^{2}_{i} }{ \left[ 1  + k_{\left(i-1\right)i} \alpha^{2}_{i} \right]^{2} },  \nonumber \\
\tilde{d}_{i} &=&  d_{i-1} \alpha^{2}_{i} +  d_{i} - d_{\left(i-1\right)i} \alpha_{i}, \nonumber \\
\alpha_{i} &=& \mp \frac{ \sqrt{  \left[ d_{i-1} k_{i\left(i-1\right)}  - d_{i} \right]^{2} + d^{2}_{\left(i-1\right)i} k_{i\left(i-1\right)} }  }{   d_{\left(i-1\right)i} }  \nonumber \\
&&- \frac{  d_{i-1} k_{i\left(i-1\right)}  - d_{i} }{   d_{\left(i-1\right)i} },
\label{eq:ChainWithNearestNeighbourInteractionEq_3}
\end{eqnarray}
where $i = 2, 4, \ldots, n$. We have therefore diagonalized the $n$-body Hamiltonian completely with all the quantities given in closed form as functions of the original parameters of the Hamiltonian, $m_{i},d_{i},d_{ij}$.

In the case of Bravais chains all the masses $m_{i}$ and the force constants $d_{i}$ are the same. We therefore denote these quantities as $m_{i} \equiv m$ and $d_{i} = d_{1}$ for all $i$. Moreover, the coefficients $d_{ij}$ have the following symmetry $d_{\left(i+k\right)\left(j + k \right)}$ for any suitable integer $k$. From this it follows that we can denote $d_{\left(i-1\right)i} = d_{i\left(i+1\right)} = d_{12}$ for all $i$. With these identifications, we write in the Bravais chain case for the quantities in Eq. \ref{eq:ChainWithNearestNeighbourInteractionEq_3}
\begin{eqnarray} 
\tilde{m}_{i-1} &=& \frac{ m }{ 2 }, \quad \tilde{m}_{i} = 2 m, \nonumber \\
\tilde{d}_{i-1} &=& \frac{ 2 d_{1} \mp d_{12} \sign d_{12} }{ 4 },  \quad \tilde{d}_{i} = 2 d_{1} \pm d_{12} \sign d_{12}, \nonumber \\
\alpha_{i} &=&  \mp \sign d_{12}, \quad i = 2, 4, \ldots, n.
\label{eq:ChainWithNearestNeighbourInteractionEq_4}
\end{eqnarray}
From these results we obtain the frequencies
\begin{eqnarray} 
\omega^{2}_{i} &=& \frac{ \Phi_{11} + \Phi_{12} }{ m }, \quad i = 2, 4, \ldots, n, \nonumber \\
\omega^{2}_{i} &=& \frac{ \Phi_{11} - \Phi_{12} }{ m }, \quad i = 1, 3, \ldots, n-1.
\label{eq:ChainWithNearestNeighbourInteractionEq_5}
\end{eqnarray}
Here we used Eq. \ref{eq:NewWayOfDiagonalizationEq_7}, assumed that $\Phi_{ij} = \Phi_{ji}$ and we have chosen the solution $\alpha_{i} = -1$ in all cases, what ever the value of $\sign d_{12}$ is. We have $n$ frequencies in total, but only two possible values for them.

The Hamiltonian given by Eq. \ref{eq:ChainWithNearestNeighbourInteractionEq_1} can be also diagonalized by using other methods like the normal coordinate transformation discussed in Sec. \ref{cha:QuadraticHamiltonians}. In the following, the Bravais chain is considered. We write the Hamiltonian in terms of the quantities $\Phi_{ij}$, apply the scale transformation $w_{i} = \sqrt{m_{i}} x_{i}$ and then write an eigenvalue equation for the matrix $D_{ij} = \Phi_{ij}/\sqrt{m_{i}m_{j}}$. The matrix $D$ is a tridiagonal Toeplitz matrix with the eigenvalues \cite{Meyer-MatrixAnalysisAndAppliedLinearAlgebra-1996}
\begin{equation}
\varpi^{2}_{s} = \frac{\Phi_{11}}{m} + 2 \frac{\Phi_{12}}{m} \cos\left(\frac{ s \pi }{ n + 1 }\right), \quad s = 1,\ldots, n.
\label{eq:ChainWithNearestNeighbourInteractionEq_6}
\end{equation}
We note that for $n = 2$, Eqs. \ref{eq:ChainWithNearestNeighbourInteractionEq_5} and \ref{eq:ChainWithNearestNeighbourInteractionEq_6} give exactly the same frequencies. The diagonal Hamiltonian is written in terms of the collective normal coordinates $q_{s}$ and is of the form given by Eq. \ref{eq:QuadraticHamiltoniansEq_5} or equivalently by Eq. \ref{eq:QuadraticHamiltoniansEq_6}. At zero temperature the total energy $E$ is the sum of zero point energies and we obtain the same result by both diagonalization methods, namely
\begin{equation} 
\sum^{n}_{s} \frac{\hbar \varpi_{s}}{2} = \sum^{n}_{i} \frac{\hbar \omega_{i}}{2} = E.
\label{eq:ChainWithNearestNeighbourInteractionEq_7}
\end{equation}

\section{Conclusions}
\label{cha:Conclusions}

We used a unitary transformation to diagonalize a generic quadratic Hamiltonian appearing in many relevant areas of physics and chemistry. As a result we obtain a diagonal Hamiltonian in the original observables, but with the effective masses and force constants replacing the original ones. The transformation works as a disentangling transformation in a sense that it connects the entangled and disentangled wave functions, both written in terms of the original position operators.

Our general methodology to diagonalize a quadratic Hamiltonian supplements the already extensive variety of approaches to establish the same task. All the approaches developed earlier for this purpose have found their place in building understanding of the wide area of physics described by the quadratic Hamiltonian. We see many potential uses for the approach presented here in these systems, including the lattice dynamics related fields of research and the study of entanglement.

\appendix

\section{Three-body case}
\label{ThreeBodyCase}

Here we consider the special case of Eq. \ref{eq:NewWayOfDiagonalizationEq_6} with $n = 3$. The number of transformations needed in the diagonalization process is $N = 3$. We establish the transformations given by Eq. \ref{eq:NewWayOfDiagonalizationEq_8} in three stages. We denote the intermediate Hamiltonians as $H' \equiv U^{\dagger}_{1} H U_{1}$, $H'' \equiv U^{\dagger}_{2} H' U_{2}$, explicitly
\begin{eqnarray} 
H' &=& \sum^{3}_{i} \frac{ p^{2}_{i} }{ 2 m'_{i} } + \sum^{3}_{i} d'_{i} u^{2}_{i} + \sum^{3}_{i<j} d'_{ij} u_{i} u_{j}, \nonumber \\
H'' &=& \sum^{3}_{i} \frac{ p^{2}_{i} }{ 2 m''_{i} } + \sum^{3}_{i} d''_{i} u^{2}_{i} + \sum^{3}_{i<j} d''_{ij} u_{i} u_{j}.
\label{eq:ThreeBodyCaseEq_1}
\end{eqnarray}
The final Hamiltonian is
\begin{equation} 
\tilde{H} = \sum^{3}_{i} \frac{ p^{2}_{i} }{ 2 \tilde{m}_{i} } + \sum^{3}_{i} \tilde{d}_{i} u^{2}_{i} + \sum^{3}_{i<j} \tilde{d}_{ij} u_{i} u_{j}.
\label{eq:ThreeBodyCaseEq_2}
\end{equation}
After the first transformation, we obtain for the quantities appearing in $H'$
\begin{eqnarray} 
m'_{1} &=& \frac{ m_{1} m_{2} }{ m_{2} + m_{1} \alpha^{2}_{1} }, \nonumber \\
m'_{2} &=& m_{2}  \left( 1 + k_{12} \alpha^{2}_{1} \right), \quad m'_{3} = m_{3}, \nonumber \\
d'_{1} &=& \frac{ d_{1} - d_{12} k_{12} \alpha_{1} + d_{2} k^{2}_{12} \alpha^{2}_{1} }{ \left( 1 + k_{12} \alpha^{2}_{1} \right)^{2} }, \nonumber \\
d'_{2} &=& d_{1} \alpha^{2}_{1} + d_{2} - d_{12} \alpha_{1}, \quad d'_{3} = d_{3}, \nonumber \\
d'_{12} &=& d_{12} - \frac{ 2 \left(  d_{1} + d_{2} k_{12} \right) \alpha_{1} }{ 1 + k_{12} \alpha^{2}_{1} }, \nonumber \\
d'_{13} &=& \frac{ d_{13} - d_{23} k_{12} \alpha_{1} }{ 1 +  k_{12} \alpha^{2}_{1} }, \quad d'_{23} = d_{23} - d_{13} \alpha_{1}.
\label{eq:ThreeBodyCaseEq_3}
\end{eqnarray}
The quantities included to $H''$ are 
\begin{eqnarray} 
m''_{1} &=& m'_{1}, \nonumber \\
m''_{2} &=& \frac{ m'_{2} m'_{3} }{ m'_{3} + m'_{2} \alpha^{2}_{2} }, \quad m''_{3} = m'_{3}  \left( 1 + k'_{23} \alpha^{2}_{2} \right), \nonumber \\
d''_{1} &=& d'_{1}, \quad d''_{2} = \frac{ d'_{2} - d'_{23} k'_{23} \alpha_{2} + d'_{3} k'^{2}_{23} \alpha^{2}_{2} }{ \left( 1 +  k'_{23} \alpha^{2}_{2} \right)^{2} }, \nonumber \\
d''_{3} &=& d'_{2} \alpha^{2}_{2} + d'_{3} - d'_{23} \alpha_{2}, \nonumber \\
d''_{12} &=& \frac{ d'_{12} - d'_{13} k'_{23} \alpha_{2} }{ 1 + k'_{23} \alpha^{2}_{2} }, \quad d''_{13} = d'_{13} - d'_{12} \alpha_{2}, \nonumber \\
d''_{23} &=& d'_{23} - \frac{ 2 \left(  d'_{2} + d'_{3} k'_{23} \right) \alpha_{2} }{ 1 + k'_{23} \alpha^{2}_{2} }.
\label{eq:ThreeBodyCaseEq_4}
\end{eqnarray}  
Finally, we transform $H''$ and obtain Eq. \ref{eq:ThreeBodyCaseEq_2} with the quantities of the following form 
\begin{eqnarray} 
\tilde{m}_{1} &=& m''_{1}  \left( 1 + k''_{31} \alpha^{2}_{3}  \right), \nonumber \\
\tilde{m}_{2} &=& m''_{2}, \quad \tilde{m}_{3} = \frac{ m''_{3} m''_{1} }{ m''_{1} + m''_{3} \alpha^{2}_{3} }, \nonumber \\
\tilde{d}_{1} &=& d''_{2} \alpha^{2}_{3} + d''_{1} - d''_{13} \alpha_{3}, \quad \tilde{d}_{2} = d''_{2}, \nonumber \\
\tilde{d}_{3} &=& \frac{ d''_{3} - d''_{13} k''_{31} \alpha_{3} + d''_{1} k''^{2}_{31} \alpha^{2}_{3} }{ \left( 1 + k''_{31} \alpha^{2}_{3} \right)^{2} },  \nonumber \\
\tilde{d}_{12} &=& d''_{12} - d''_{23} \alpha_{3}, \quad \tilde{d}_{13} = d''_{13} - \frac{ 2 \left(  d''_{3} + d''_{1}  k''_{31} \right) \alpha_{3} }{ 1 + k''_{31} \alpha^{2}_{3} }, \nonumber \\
\tilde{d}_{23} &=& \frac{ d''_{23} - d''_{12} k''_{31} \alpha_{3} }{ 1 +  k''_{31} \alpha^{2}_{3} }.
\label{eq:ThreeBodyCaseEq_5}
\end{eqnarray}
The Hamiltonian $\tilde{H}$ is diagonal if
\begin{eqnarray} 
0 &=& d''_{12} - d''_{23} \alpha_{3}, \nonumber \\
0 &=& d''_{13} - \frac{ 2 \left(  d''_{3} + d''_{1} k''_{31} \right) \alpha_{3} }{ 1 + \alpha^{2}_{3} k''_{31} }, \nonumber \\
0 &=& \frac{ d''_{23} - d''_{12} \alpha_{3} k''_{31} }{ 1 + \alpha^{2}_{3} k''_{31} } .
\label{eq:ThreeBodyCaseEq_6}
\end{eqnarray}
We can solve the quantities $\alpha_{i}$ ($i = 1,2,3$) from this set of three equations as a function of the original masses $m_{i}$ and coupling constants $d_{i}$, $d_{ij}$. Given Eq. \ref{eq:ThreeBodyCaseEq_6} holds, we have obtained a diagonal Hamiltonian $\tilde{H}$ given by Eq. \ref{eq:NewWayOfDiagonalizationEq_12} such that the quantities $\tilde{m}_{i}$ and $\tilde{d}_{i}$ are given by Eq. \ref{eq:ThreeBodyCaseEq_5}. These results can be used, for instance, to solve the $n$-body case with the second nearest neighbour interactions. We leave the more detailed analysis of the 3-body case to future work.


%\begin{acknowledgments}
%V. J. H. thanks Dr. Ivan Gonoskov for useful discussions on various aspects of the present work.
%The author gratefully acknowledge funding from the Foundation for Research of Natural Resources in Finland (grant 17591/13). T
%\end{acknowledgments}
%\bibliographystyle{plain}
\bibliography{bibfile}
\end{document}